\documentclass[apj,graphicx]{emulateapj}
\usepackage{apjfonts}
\usepackage{lscape}

\begin{document}

\title{A search for iron emission lines in the 
{\it Chandra} X-ray spectra of neutron star low-mass X-ray binaries}
\shortauthors{Cackett et al.}
\shorttitle{Iron lines in neutron star low-mass X-ray binaries}

\author{E.~M.~Cackett\altaffilmark{1,2}, 
J.~M.~Miller\altaffilmark{1},
J. Homan\altaffilmark{3},
M.~van~der~Klis\altaffilmark{4},
W.~H.~G.~Lewin\altaffilmark{3},\\
M.~M\'endez\altaffilmark{5},
J.~Raymond\altaffilmark{6},
D.~Steeghs\altaffilmark{6,7},
R.~Wijnands\altaffilmark{4}}
\email{ecackett@umich.edu}
\altaffiltext{1}{Department of Astronomy, University of Michigan, 500 Church St, Ann Arbor, MI 48109-1042, USA}
\altaffiltext{2}{Dean McLaughlin Postdoctoral Fellow}
\altaffiltext{3}{MIT Kavli Institute for Astrophysics and Space Research, MIT, 70 Vassar Street, Cambridge, MA 02139-4307, USA}
\altaffiltext{4}{Astronomical Institute `Anton Pannekoek', University of Amsterdam, Kruislaan 403, 1098 SJ, Amsterdam, the Netherlands}
\altaffiltext{5}{Kapteyn Astronomical Institute, University of Groningen, P.O. Box 800, 9700 AV Groningen, the Netherlands}
\altaffiltext{6}{Harvard-Smithsonian Center for Astrophysics, 60 Garden Street, Cambridge, MA 02138, USA}
\altaffiltext{7}{Department of Physics, University of Warwick, CV4 7AL, UK}

\begin{abstract}
While iron emission lines are well studied in black hole systems, both in X-ray binaries and Active Galactic Nuclei, there has been less of a focus on these lines in neutron star low-mass X-ray binaries (LMXBs).  However, recent observations with {\it Suzaku} and {\it XMM-Newton} have revealed broad asymmetric iron line profiles in 4 neutron star LMXBs, confirming an inner disk origin for these lines in neutron star systems.  Here, we present a search for iron lines in 6 neutron star LMXBs.  For each object we have simultaneous {\it Chandra} and {\it RXTE} observations at 2 separate epochs, allowing for both a high resolution spectrum, as well as broadband spectral coverage.  Out of the six objects in the survey, we only find significant iron lines in two of the objects, GX~17+2 and GX~349+2. However, we cannot rule out that there are weak, broad lines present in the other sources.  The equivalent width of the line in GX~17+2 is consistent between the 2 epochs, while in GX~349+2 the line equivalent width increases by a factor of $\sim$3 between epochs as the source flux decreases by a factor of 1.3.  This suggests that the disk is highly ionized, and the line is dominated by recombination emission.  We find that there appears to be no specific locations in the long-term hardness-intensity diagrams where iron emission lines are formed, though more sources and further observations are required.
\end{abstract}

\keywords{accretion, accretion disks --- stars: neutron --- X-rays: binaries}

\section{Introduction}
Accretion disks around compact objects in AGN and X-ray binaries exist in strong gravity around these objects, and thus can provide a means to study gravity in the strong-field regime.  In many of these objects, an Fe K-shell fluorescence emission line is observed in the X-ray spectra \citep[e.g.][]{white86,nandra97,reynolds97,asai00,miller04}.  These lines are formed in the inner accretion disk and are observed to be broadened due to strong Doppler shifts and gravitational redshifts in that region \citep[see][for a review of iron lines]{miller07}.  As these lines are shaped by the motions of gas in the disk, they are an important diagnostic of the inner accretion disk radius in these objects.

Observations of black holes (in both AGN and X-ray binaries) have revealed some of these lines to be highly asymmetric \citep[e.g.][]{tanaka95,fabian02,miller04} due to the extreme motions of gas in the inner disk around black holes.  While iron lines in black holes are now well-studied, iron lines in neutron star low-mass X-ray binaries (LMXBs) have not been studied in the same detail.  Nevertheless, iron lines in neutron star LMXBs are well known, and these lines are significantly weaker than in black holes \citep[e.g.][]{white85,white86,hirano87,asai00,disalvo05}. Thus, the line shapes could not be studied in detail, and could be modeled as just a Gaussian.  Only with recent sensitive {\it XMM-Newton} and {\it Suzaku} observations have broad, asymmetric iron lines been observed in neutron stars, and these observations highlight the potential to use iron lines in neutron stars as a probe of the inner disk radius; the inner disk radius is an upper limit on the neutron star radius \citep{bhattacharyya07,cackett08}.

The use of these lines as a disk diagnostic also allows for a test of the standard neutron star accretion picture described by color-color diagrams and kilo-Hertz quasi-periodic oscillations (kHz QPOs).  These kHz QPOs are seen to vary in frequency as the neutron star changes state and moves around the color-color diagram \citep[see][for a comprehensive review]{vanderklis06}.  The kHz QPOs have frequencies that may be associated with the orbital frequency of the inner disk and there are several proposed mechanisms for their production \citep[e.g.][]{miller98,stella99,lamb01,abramowicz03}.  By combining color-color diagrams, QPOs and iron lines, it may be possible to break the degeneracies in the models.  For instance, if they are due to orbital motions in the disk, then a change kHz QPO frequency may correspond to a change in the inner accretion disk radius.  However, in disk resonance models this would not be expected.  If the inner disk radius can be measured independently using iron lines, there is the potential to test these models.  In fact, recent observations of iron lines in 3 systems (Ser~X$-$1, GX~349+2 and 4U~1820$-$30) show a good agreement between the measured inner disk radius from the iron lines and the inferred inner disk radius from kHz QPOs \citep[when the highest observed kHz QPO frequency is used;][]{cackett08}. It is therefore important to identify as many neutron stars with iron lines as possible, to determine whether such a method is possible.  The location on hardness-intensity (or alternatively color-color) diagrams, where iron lines are present, may also be key, and in particular how this corresponds to where kHz QPOs are seen.

Moreover, iron lines in neutron star LMXBs can provide a test of iron line models for black holes.  In several objects extremely broad lines are observed from black holes.  It has been proposed that the extreme width of the lines requires that the black holes are spinning \citep[e.g.][]{iwasawa96,wilms01,fabian02,miller_xtej1650_02,miller04}, and thus allowing the disk to extend closer in than the inner-most stable orbit for a non-spinning black hole.  Of course, in neutron stars the solid surface of the star prevents the accretion disk from extending any further in than that.  Any iron lines in neutron star LMXBs should therefore be significantly less broad than those in spinning black holes.  Results from recent sensitive observations do find that the neutron star lines are narrower than in the proposed spinning black holes, supporting that hypothesis \citep{bhattacharyya07,cackett08}.

As part of the {\it Chandra} HETGS Z/Atoll Spectroscopic Survey \citep[CHAZSS, see][for related work]{cackett1820} we have obtained observations of 6 neutron star LMXBs.  Each source in this survey was observed at 2 separate epochs using the {\it Chandra} High Energy Transmission Grating Spectrometer (HETGS).  Simultaneously, we observed these sources with the {\it Rossi X-ray Timing Explorer (RXTE)}.  In this paper we present the search for Fe K$\alpha$ emission lines in these sources and we examine how the presence, or lack of iron lines corresponds to position in hardness-intensity diagrams.  In Section \ref{sec:data} we describe the observations and data reduction.  Section \ref{sec:results} details our analysis and results, and in Section \ref{sec:disc} we discuss the implications of the iron lines that we find.

\section{Observations and Data Reduction} \label{sec:data}
Persistent neutron star LMXBs are frequently referred to as being either atoll or Z sources due to the shape of the tracks they trace on 
X-ray color-color and hardness-intensity diagrams \citep{hasinger89}. The differences between the two types of sources is thought to be due to different mass accretion rates, with the Z sources accreting at higher fractions of the Eddington rate than atolls \citep[see][for a review]{vanderklis06}.  This picture was recently strengthened by observations of the first transient Z source XTE~J1701$-$462 \citep{homan07} whose behavior became more like atolls as it decreased in luminosity. 

Six neutron star LMXBs are part of CHAZSS.  Of these sources three are classified as atolls (4U~1636$-$53, 4U~1735$-$44, 4U~1820$-$30) and three are classified as Z sources (GX~17+2, GX~340+0 and GX~349+2). We observed each of these objects twice with {\it Chandra} and simultaneously with {\it RXTE}.  Details of these observations are given in Table \ref{tab:obs}.  The {\it Chandra} and {\it RXTE} data reduction is described in the following sections.

\subsection{Chandra data reduction}
During each observation the HETGS spectrum was read-out with the ACIS-S array operating in continuous-clocking (CC) mode. Even though these sources are bright, the 3 msec frame time in CC mode means that in the dispersed spectrum the count rate per pixel is not high enough to cause pile-up. 
A gray filter 100 columns wide was used over the zeroth order.  This prevents telemetry saturation yet still leaves the position of the zeroth order visible, which is vital for accurately extracting spectra.

We performed the data reduction for each observation in the same manner using CIAO version 3.3.0.1 and following the standard analysis threads\footnote{http://cxc.harvard.edu/ciao/threads}.  
The data was reprocessed from the level 1 event file to use the latest calibrations.  The location of the zeroth order was determined using the {\it tgdetect} tool.  Diffracted grating events were assigned to spectral orders (using {\it tg\_resolve\_events}), with the spectral orders separated using the intrinsic energy resolution of the detectors.  The data was then filtered for bad grades and for a clean status column (status=0) before applying the good time interval filters.  Finally, before extracting the spectra the {\it destreak} tool was used to remove streaking effects on the ACIS-S4 chip.

The {\it tgextract} tool was used to extract the +1 and -1 order spectra for the High Energy Grating (HEG), with the nominal instrument resolution.  We used the standard redistribution matrix files (RMFs) from the CALDB, and generated the ancillary response files (ARFs) using the {\it fullgarf} script.  The HEG +1 and -1 orders were examined to ensure that there was no wavelength shift between the orders before the spectra and ARFs were combined using the {\it add\_grating\_spectra} script, giving a combined first-order spectrum for the HEG.  We do not use the Medium Energy Grating (MEG) data here as the effective area drops off significantly over the Fe K band (6.4-6.97 keV).

\subsection{RXTE data reduction}

The {\it RXTE} observations we present here are from pointed observations using the proportional counter array (PCA).  The PCA data were reduced using the LHEASOFT suite (version 6.1).  We performed the standard `goodtime' filtering and `deadtime'  corrections using data when the pointing offset $< 0.02^{\circ}$ and earth-limb elevation angle was larger than $10^{\circ}$.  We use data from PCU 2 only as this is the most reliable of all five PCUs.   Spectra were extracted from the `Standard 2' mode data, with systematic errors of 0.6\% applied to each channel.  We used the appropriate background model depending on whether the source was faint or bright (with faint defined as a source count rate of $< 40$ counts/s/PCU).  Background subtracted lightcurves were also created from the `Standard 2' mode data.  X-ray bursts were searched for in the lightcurves of all observations but were only seen in 4U~1636$-$53.  We remove all data between 20 s before and 200 s after the X-ray bursts.

In order to study the spectral state of the sources we chose to create hardness-intensity diagrams for each source.  For this purpose, we define the hard color as the ratio of count rates in the $9.7-16.0$ keV and $6.4-9.7$ keV bands.  We extracted lightcurves (with 256s binning for the atoll sources and 128s binning for the Z sources) in each of these energy bands to produce hard color curves that we normalized using simulated Crab-like spectra, to account for long term changes in the PCA response. These Crab-like spectra were produced with the ``fakeit" command in XSPEC, using response matrices that were created by PCARSP V10.1. We also extracted the source intensities across the $3.0-25.0$ keV band, which are also normalized in the same way.

As we want to compare each source with its position in the long-term hardness-intensity diagram, we obtained every {\it RXTE}/PCA observation of each source in our survey that was publically available in the archive before January 2007.  We reduced the data in the same manner as above to produce the long term hardness-intensity diagrams.

\section{Analysis and Results} \label{sec:results}

\subsection{Spectral Fitting}
In observing iron lines it is important to have a well-defined continuum on either side of the Fe K band ($6.4-6.97$ keV). We therefore fitted both the {\it Chandra} HEG and {\it RXTE} PCA spectra simultaneously.  The addition of the {\it RXTE} PCA spectra provides a larger energy range above the Fe K band where the continuum is well-defined (as opposed to fitting the {\it Chandra} data alone), and thus allows for a more accurate determination of the model components necessary to describe the continuum emission, facilitating the search for iron lines.

For a long time, there has been ambiguity as to the correct spectral model to fit to neutron star LMXBs, with a variety of different spectral models all fitting equally well \citep[e.g.][]{white88,mitsuda89}. Recent work by \citet*{lin07} explored the different models by studying how the measured temperature of the thermal components varied over a wide range in luminosity. This was done using extensive {\it RXTE} observations of two transient atoll sources Aql~X$-$1 and 4U~1608$-$52.  Their preferred spectral model was determined from the model where the measured temperature best followed $L \propto T^4$, and differs for the hard state and the soft and transitional states.  For the hard state, they find that a single temperature blackbody plus a broken power-law performs best, with the broken power-law used to mimic weak Comptonization.  For the soft and transitional states, their preferred model has two thermal components (a blackbody and a multi-color disk blackbody) representing the boundary layer and accretion disk, plus a constrained broken power-law to model Comptonization.  The constrained broken power-law has the break energy fixed at 20 keV, and the first power-law index constrained to be less than 2.5.

\subsubsection{Hard state observation}

Here, we adopt a very similar prescription to \citet{lin07}.  Of all the observations,we find that only one is in the hard state (the first observation of 4U~1636$-$53).  Spectral fitting was performed using XSPEC v11 \citep{arnaud96}.  For this hard state observation we modeled the continuum using a single-temperature blackbody (\verb|bbody| in XSPEC) plus a broken power-law (\verb|bknpower| in XSPEC), absorption from the interstellar medium was accounted for using the model \verb|phabs|.  We fitted the {\it Chandra} HEG spectra in the $2-9$ keV range and the {\it RXTE} PCA spectra between 3.6 and 25 keV.  In the {\it RXTE}/PCA spectrum above 25 keV the background dominates over the source flux, we therefore do not fit any energies higher than this.  Below 2 keV we find that the {\it Chandra} HEG spectrum turns up slightly, regardless of the model being used.  We attribute this to scattered light off the telescope support structure which is not quite corrected for with the current calibration.

When performing the spectral fitting, we tie all the continuum parameters between the {\it Chandra} and {\it RXTE} spectra except the absorbing column density, which was allowed to differ.  We found that this significantly improved the fit, helping to correct for the scattered light.  Importantly, this does not affect the fit through the Fe K band and at higher energies.  A constant is also included in the model to allow for differences in absolute flux calibration between the two instruments.  Moreover, we found that a broad iron line was present in the {\it RXTE}/PCA data, thus we added in a Gaussian component (with a central energy constrained to be within $6.4-6.97$ keV) to account for this.  However, it is the properties of any iron line present in the {\it Chandra} data that is of interest here -- the significantly higher spectral resolution of the {\it Chandra} gratings is required to study the iron line properties in detail.  The spectral fitting results are given in Table \ref{tab:spectra_hard}.  The fluxes given are from fits to the {\it Chandra} data, with the 10-20 keV flux determined from an extrapolation of the best-fitting model to higher energies.

\subsubsection{Soft/Transitional state observations}

All other observations are in the soft/transitional states, thus we use a blackbody (\verb|bbody|) plus a disk blackbody (\verb|diskbb|) model.  When statistically required we also include a power-law component, constrained to have a spectral index less than 2.5.  Again, we fitted the {\it Chandra} HEG spectra in the $2-9$ keV range, however, for the {\it RXTE} PCA spectra, we only fit from 3.6 to 20 keV.  This allows us to simplify the \citet{lin07} prescription - their model has a constrained broken power-law with the break energy fixed at 20 keV.  By ignoring data above 20 keV we can simplify the model by just using a single unbroken power-law without affecting the fit at lower energies.  For the two brightest sources (GX~17+2 and GX~349+2) we found that the scattered light effect was worst, and hence we had to restrict the energy range of the {\it Chandra} HEG to $3-9$ keV for those spectra only.

Again, we included a constant to account for differences in the absolute flux calibration between the instruments.  As with the hard state observation, all continuum parameters were tied between the {\it Chandra} and {\it RXTE} spectra except the absorbing column, which was allowed to vary (for the same reasons mentioned above).  In some cases where the column density is not well-constrained and tends to zero in the {\it Chandra} spectra, we fix the column density at the Galactic value in that direction, while still allowing it to be a free parameter in the {\it RXTE} data.   A Gaussian component was used to model the broad iron line seen in all {\it RXTE} spectra apart from GX~340+0, where one was not required.  We stress again that the limited spectral resolution of {\it RXTE}/PCA hampers the use of that data to study the iron line properties in detail.

\begin{figure}
\centering
\includegraphics[width=14cm]{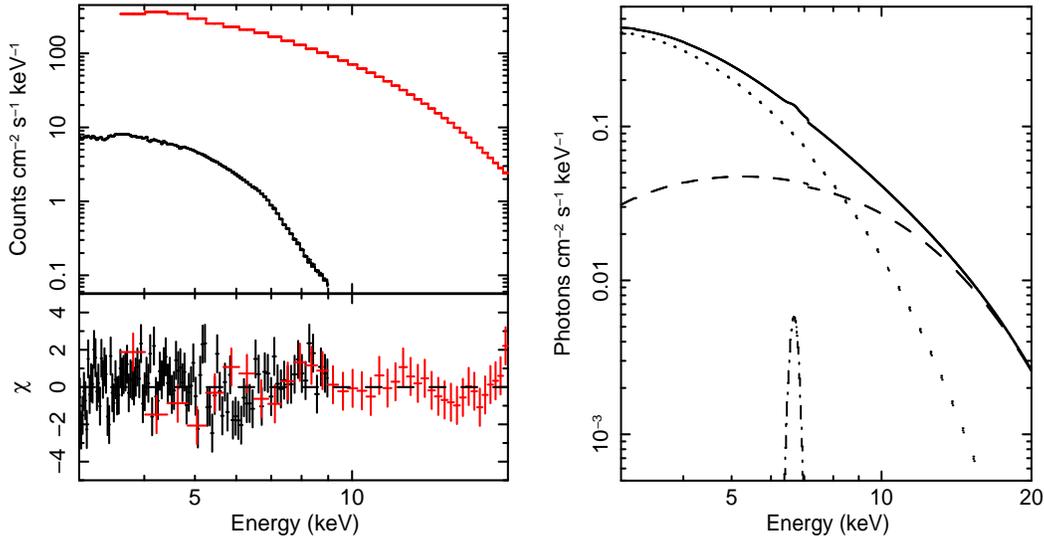}
\caption{{\it Left:} Chandra (black) and RXTE/PCA (red) spectra from the first observation of GX~17+2.  The bottom panel shows the residuals ($\chi$ = (data-model)/$\sigma$).  {\it Right:}  The best fitting model components to the {\it Chandra} data are shown (extended to show the model up to 20 keV).  The solid line is the overall model spectrum, the dotted line is the disk blackbody component, the dashed line is the blackbody component and the dash-dot line is the Gaussian modeling the iron line.}
\label{fig:spec_mod}
\end{figure}

We found that for all the atoll sources in our sample (4U~1636$-$53, 4U~1735$-$44, and 4U~1820$-$30) the constrained power-law component was statistically required.  However, for the Z sources (GX~17+2, GX~340+0 and GX~349+2) the constrained power-law did not significantly improve the quality of the fit, and thus was not used.  The spectral fitting results are given in Table \ref{tab:spectra}, and we show the {\it Chandra} HEG and {\it RXTE} PCA spectra for the first observation of GX~17+2 in Fig.~\ref{fig:spec_mod} along with the best-fitting model components to the {\it Chandra} spectrum.

\subsubsection{Searching for iron emission lines in the {\it Chandra} spectra}

\begin{figure}
\centering
\includegraphics[width=14cm]{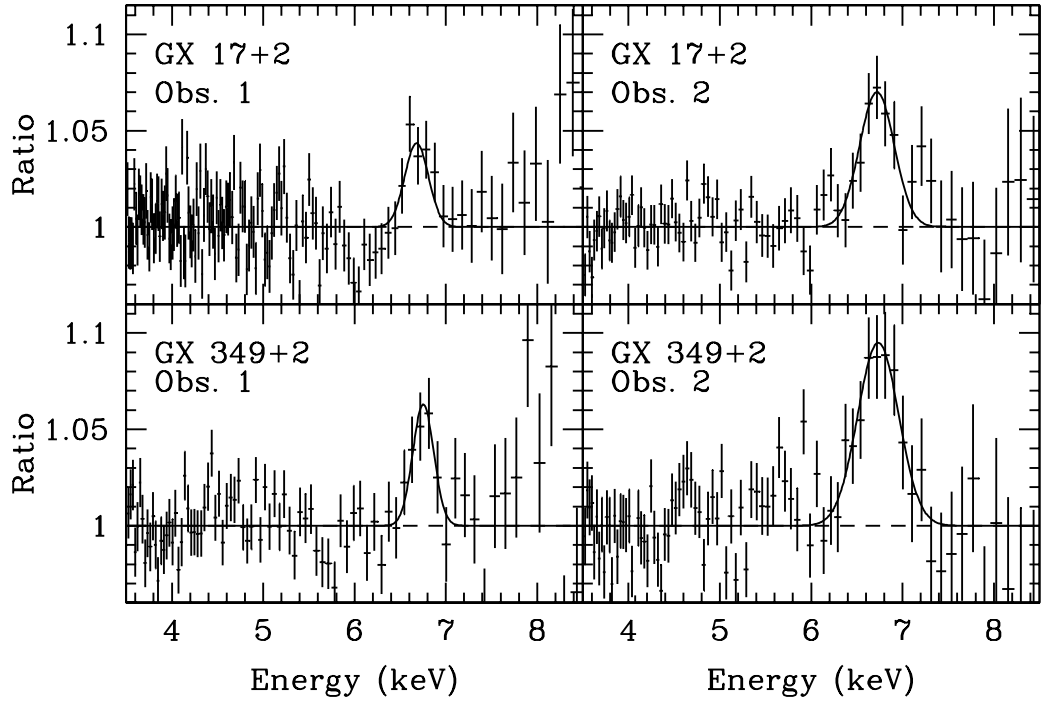}
\caption{Iron emission lines detected in the {\it Chandra} HEG spectra of GX 17+2 (top) and GX 349+2 (bottom).  The first observations are on the right, and the second observations are on the left.  Plotted is the ratio of the data to the continuum model.  The solid line is the best-fitting continuum + line model.}
\label{fig:lines}
\end{figure}

Having determined the model components needed to fit the continuum, we look for iron emission lines in the {\it Chandra} HEG spectra.  This was performed by fitting a Gaussian to the {\it Chandra} HEG spectrum with the energy constrained to be within the $6.4-6.97$ keV Fe K range, and determining whether this additional model component improves the fit significantly. Note that after adding in the Gaussian component to the model, we still allow the continuum parameters to vary in the fit.  We detect significant iron lines in the {\it Chandra} HEG spectra of two objects: GX~17+2 and GX~349+2.  In these objects an iron line is present in both observations.  Significant iron lines are not observed in the {\it Chandra} spectra of the other four objects.  We estimate the line significance from the uncertainty in the equivalent width.  The detected iron lines are shown in Fig. \ref{fig:lines}, and the parameters from the Gaussian fits are given in Table \ref{tab:lines}.

We put upper limits on lines in the other spectra by fitting a Gaussian with a fixed width of $\sigma=0.1$ keV and also $\sigma=0.5$ keV.  We choose these line widths to probe for both narrow and broad lines. The energy of the line was allowed to be a free parameter, apart from those cases where it could not be constrained within $6.4-6.97$ keV.  In those cases, the line energy was fixed to 6.7 keV, consistent with the detected lines.  The determined upper limits are given in Table \ref{tab:nonsig_lines}. 

\begin{figure}
\centering
\includegraphics[width=8.4cm]{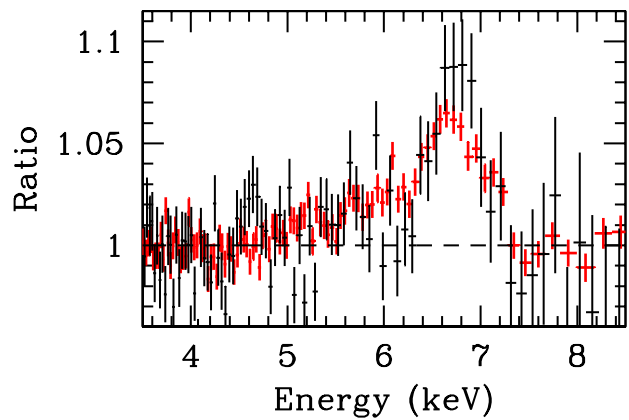}
\caption{A comparison of the iron emission lines detected from GX 349+2 with {\it Chandra} (black, 2nd observation) and {\it Suzaku} (red, only XIS detector 2 is shown).}
\label{fig:cxo_suz}
\end{figure}

The iron lines observed in GX~17+2 and GX~349+2 do not appear to be asymmetric.  This is due to the low signal to noise in the continuum redward of the line, as {\it Suzaku} observations of the line in GX 349+2 clearly show it as being asymmetric \citep{cackett08}.  A simple Gaussian fits the lines well in all observations.  While the more complex relativistic accretion disk emission line model, \verb|diskline| \citep{fabian89}, can also fit the lines well, when all the parameters are left free in the fit they are not well constrained and the model does not statistically improve the fit over a Gaussian.  To demonstrate that an asymmetric line cannot be ruled out we show both the {\it Chandra} and {\it Suzaku} data together in Fig. \ref{fig:cxo_suz}. The inner disk radius determined from fitting the {\it Chandra} GX~349+2 data with the \verb|diskline| model can be significantly improved by using the emissivity index, $\beta$ and disk inclination, $i$, found from fits to the {\it Suzaku} data in \citet{cackett08}. We therefore fitted the 2nd {\it Chandra} observation (where the iron line is the strongest) with the \verb|diskline| model, fixing $\beta = -4.1$, $i = 23^{\circ}$ and constraining the line energy to be within $6.4-6.97$ keV.  Doing this, we measured the inner disk radius, $R_{in} = 10.45\pm0.49~GM/c^2$, normalization = $(1.1\pm0.4)\times10^{-2}$, and equivalent width = $92\pm31$ eV.  The equivalent width is larger than fitting a Gaussian due to the asymmetric profile of the \verb|diskline| model.

The equivalent width of the lines observed in GX~349+2 are smaller than those observed with {\it Suzaku} \citep{cackett08}, where they find EW = $76\pm6$ eV.  However, this difference can be attributed to the asymmetric line profile given that fitting the 2nd observation with a \verb|diskline| model gives a larger equivalent width than fitting with a Gaussian, and one that is consistent with the {\it Suzaku} value.  We note the first observation is consistent with the Gaussian fit to the line in this object by \citet{iaria04}, who find $39\pm13$ eV.  The equivalent widths measured for GX~17+2 are consistent with previous observations of the iron line in this source \citep{farinelli05,disalvo00}.

Comparing the equivalent widths of the detected lines between each epoch we find that the lines in GX~17+2 are almost consistent with remaining unchanged between the observations.  On the other hand in GX~349+2 the equivalent width is larger by a factor of $\sim$3 in the second observation, though this increase is of low significance.  One might expect that the line equivalent width would increase with increasing hard X-ray flux (with energies greater than the line energy).  However, we find that the $10-20$ keV flux decreases between observations, the opposite of the change in equivalent width.  The $2-10$ keV flux also decreases between observations by a similar amount, thus the overall spectral shape is not very different.

Of the sources where we do not detect iron lines in the {\it Chandra} spectra, only 4U~1820$-$30 has had a previously detected iron line with CCD-resolution spectroscopy.  Recent {\it Suzaku} observations detected a broad line in 4U 1820-30 \citep{cackett08}.  Moreover, a line was also observed in the object previously with {\it ASCA}  \citep{asai00}. Nevertheless, we do not detect any line here.  However, the upper limits we determine with $\sigma=0.5$ keV are not too far from the  equivalent width as measured in recent {\it Suzaku} observations by \citet{cackett08}, where EW $= 51\pm11$ eV.  This indicates that the lower signal to noise here may prevent a significant detection of the line.  We therefore cannot exclude that weak, broad lines may also be present in the other sources as the observations are not sensitive enough to detect them.  In fact, we note that in 4U~1636$-$53 there is evidence of a broad iron line in recent {\it Suzaku} observations (Homan et al, private communication), as well as from {\it XMM-Newton} observations \citep{pandel08}.

\subsection{Hardness-intensity diagrams and timing analysis}

\begin{figure}
\centering
\includegraphics[width=14cm]{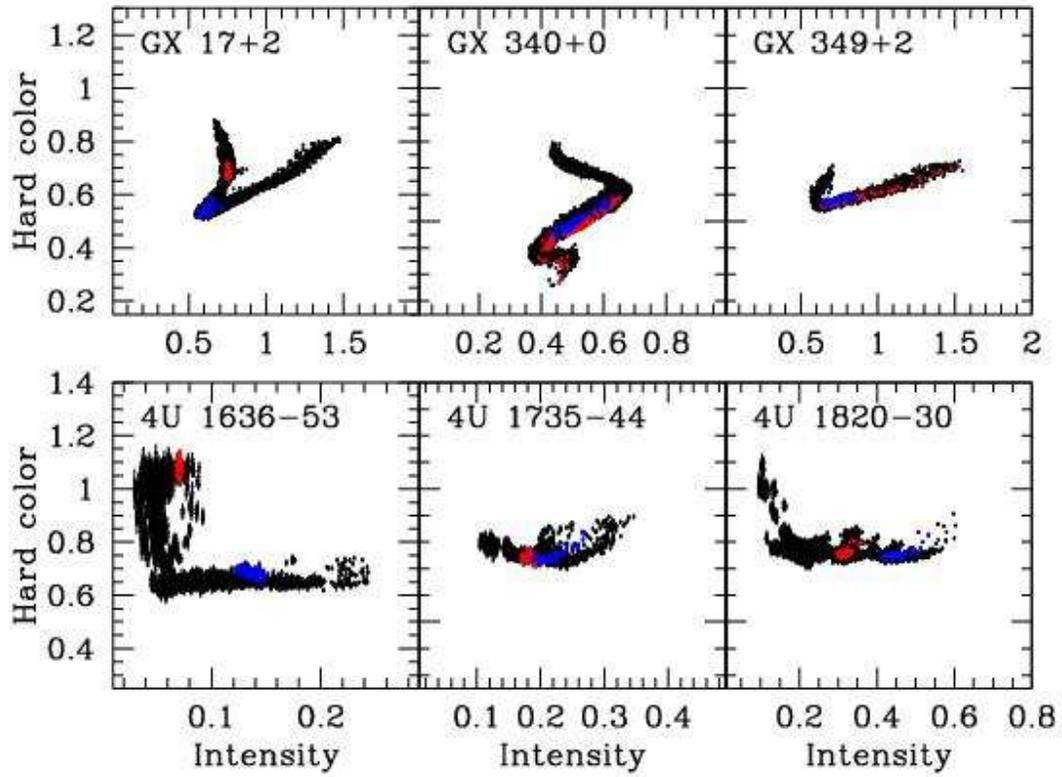}
\caption{Long-term {\it RXTE} hardness-intensity diagrams for all six sources.  The first CHAZSS observations are red and the second CHAZSS observations are blue.  The intensities are normalized to the Crab in the 3.0-25 keV band.}
\label{fig:colint}
\end{figure}

The location of the CHAZSS observations on the long-term {\it RXTE} hardness-intensity diagrams is shown in Fig. \ref{fig:colint}.  The first CHAZSS observation is in red, the second CHAZSS observation in blue and the other observations in black.  As discussed above, significant iron emission lines are observed in the {\it Chandra} spectra of GX~17+2 and GX~349+2, but not in any of the other sources.  Comparing the hardness-intensity diagrams it appears that there may be no special location where iron emission lines are preferentially observed (though observations in more states are needed to support this).  The implications of this are discussed in the next section.

Of our observations of the two sources that show iron lines, only the first observation of GX~17+2 is in the correct state to display kHz QPOs (as indicated by its location on the hardness-intensity diagram).  To investigate this further we extracted fast timing mode data from the 3.6 -- 60 keV band using all available PCUs (which ranged from between 2 and 4 during the observations).  We created a broadband power density spectrum using the standard fast Fourier transform techniques \citep{vanderklis89,vanderklis95}.  For the first GX 17+2 observation we find that there is a pair of kHz QPOs present in the power spectrum, with frequencies of $594\pm11$ Hz and $906\pm15$ Hz, and fractional rms amplitudes of $2.4\pm0.3\%$ and $3.3\pm0.4\%$ respectively.  The quality factor which is defined as $Q = \nu_0/2\Delta$ (where $\nu_0$ is the centroid frequency and $\Delta$ the half-width at half-maximum) is $5.1\pm1.2$ for the lower frequency kHz QPO and $3.5\pm1.0$ for the upper kHz QPO.  We find there are no kHz QPOs in the second observation of GX~17+2.  We also analyzed the fast timing mode data for both GX~349+2 observations in the same manner and found that no kHz QPOs were present, as expected based on the sources position on the hardness-intensity diagram during these observations.

\section{Discussion} \label{sec:disc}
We present {\it Chandra} and {\it RXTE} observations of six neutron star low-mass X-ray binaries at two separate epochs (4U~1636$-$53, 4U~1735$-$44, 4U~1820$-$30, GX~17+2, GX~340+0 and GX~349+2).  We detect iron emission lines in the {\it Chandra} HEG spectra of both GX~17+2 and GX~349+2 at both epochs.  No iron lines are detected in the {\it Chandra} spectra of the other four sources, although we do not have the sensitivity to rule out that weak, broad lines are present.  Comparing the position of the sources in hardness-intensity diagrams (using the {\it RXTE} data) we find that from these observations there appears to be no region of the diagram where the lines are preferentially observed. However, we note that we only sample a small number of states here, and more sensitive observations are needed to confirm this. Such a finding is interesting though since we expect that the hardness-intensity diagrams are related to the mass accretion rate and the accretion disk-corona interaction. For example, changes from bright, soft (disk dominated) states to fainter, hard (corona dominated) states may indicate changes in mass accretion rate and the prominence of the disk or the corona.  One might expect to see changes in the iron lines as a function of the prominence of the disk or corona. 

The 3 Z sources observed are seen across most locations in the color-color and color-intensity diagrams.  GX~17+2 is seen at both the normal branch (NB)/horizontal branch (HB) transition as well as the NB/flaring branch (FB) transition. GX~340+0 is seen mainly on the NB, with part of the first observation on the FB. GX~349+2 is observed on the FB in both observations.  Thus, iron lines are detected in the {\it Chandra} spectra on both the FB, and at the NB/HB and NB/FB transitions. 

Previous {\it BeppoSAX} observations of GX~17+2 and GX~349+2 \citep{disalvo00,disalvo01} support our finding that iron lines are seen in no special location on the hardness-intensity (or color-color) diagram.  During the {\it BeppoSAX} observations of GX~17+2 the source was in the horizontal and normal branches.  When the data was split based on state, an iron line with an equivalent width in the approximate range $30-40$ eV was observed in all spectra \citep{disalvo00}.  For the {\it BeppoSAX} GX~349+2 observations the source was observed on the flaring branch, and both non-flaring and flaring spectra showed an iron line \citep{disalvo01}.

We do not see any iron lines in the {\it Chandra} spectra in any of the atoll sources in our sample.  The majority of the observations are spread out over the softest part of the color-color diagram - the banana branch, although we do observe 4U~1636$-$53 in the hard (island) state.  This indicates that we do not observe strong lines in either spectrally soft or hard states in the atolls.  However, we cannot rule out that weak, broad lines are present in the spectra as has been seen with more sensitive observations of 4U~1820$-$30 \citep{cackett08}. In fact, while we don't detect iron lines during the banana branch here, observations of other neutron star LMXBs, such as the atoll 4U~1705$-$44, do show iron lines during the banana branch \citep{disalvo05,piraino07}, supporting that iron lines can be seen in a wide variety of spectral states.

One interesting proposition is that of using sensitive observations of iron lines to help constrain models for kHz QPOs.  Additionally one can combine the measurement of both the disk velocity (from iron lines) and orbital frequency (from kHz QPOs, assuming that is their origin) to determine the neutron star mass \citep{cackett08}.  For this to be possible we would need to see both iron lines and kHz QPOs simultaneously.  Here we have detected both an iron line and twin kHz QPOs in the case of GX~17+2. However, the observation of GX~17+2 is not sensitive enough to detect any broad red wing of the iron line, and thus we cannot get a reliable measure of the inner accretion disk radius by modeling the iron line.  Nevertheless, this is promising for further studies using more sensitive iron line observations, such as can be obtained with {\it Suzaka}.

There are several possibilities for the source of ionizing flux that gives rise to the iron emission lines.  It may be irradiation by a corona, by the base of a compact jet, or by the boundary layer.  These observations allow us to investigate these possibilities.  For instance, the hardest states in the hardness-intensity diagram are usually interpreted as being the state where the corona is most dominant, thus we may expect to have stronger iron lines in harder spectral states.  Or, if the ionization is too high in the hardest states, the iron lines may get stronger as the source gets softer, but would be expected to weaken or even disappear in the disk dominated states.  In addition, the irradiation may be due to the boundary layer, or the base of a compact jet.  If the irradiation is due to the base of a compact jet, then we would expect to see iron lines when the jet is present, and not when the jet disappears.

We find that the iron lines detected in GX~17+2 and GX~349+2 are both strongest when the sources are closest to the NB/FB transition (though the changes in equivalent width are not highly significant). This is similar to what was found by \citet{disalvo00, disalvo01} for these sources. Comparing the 10-20 keV flux with the line equivalent widths, we find that in both GX~17+2 and GX~349+2 the equivalent width increases with decreasing 10-20 keV flux.  This can be explained if the gas in the disk is very highly ionized already and thus the emission is mostly due to recombination.  In such a scenario a drop in hard flux should lower the ionization leading to more recombinations and a stronger line.  The fact that we see the line energies consistent with the He-like 6.7 keV iron line is in agreement with this.  The absence of strong lines in sources with short periods (such as 4U~1820$-$30) is also consistent with such a picture, as the disks should be highly ionized in these sources.

X-ray and radio observations of neutron star LMXBs have associated jet formation and location on the color-color/hardness-intensity diagrams \citep[e.g.,][]{penninx88,hjellming90a,hjellming90b,migliari06,migliari07}.  From a detailed study of GX~17+2, \citet{migliari07} find that radio emission is not detected in the FB and turns on as the source transitions into the NB.  The radio flux density increases through the NB and is strongest during the HB.  We have concurrent radio observations of the sources in this study with the VLA (Cackett et al., in preparation), and thus can look for any correlations between radio flux density and the iron lines.  For GX~349+2 we do not detect the source at 5 GHz in either observation (consistent with its position on the FB), yet for GX~17+2 we have flux densities of $1.80\pm0.05$ mJy and $0.9\pm0.07$ mJy at 5 GHz for observations concurrent with observations 1 and 2, respectively.  Thus, in GX~17+2 we see higher radio flux densities at the NB/HB state than in the NB/FB state \citep[consistent with the picture of][]{migliari07}.  Note though that while there is a significant change in radio flux between observations the equivalent width of the iron line in GX~17+2 does not change significantly.  In addition, given that we do not detect any radio emission from GX~349+2 yet we do see an iron line this would tend not to favor a compact radio jet as the source of ionizing flux for the line emission.  However, it is not clear whether a disk corona or the boundary layer is the more likely source of irradiation.

\acknowledgements

JMM gratefully acknowledges support from {\it Chandra}.

\bibliographystyle{apj}
\bibliography{apj-jour,felines}

\begin{deluxetable}{cccccc}
\tablecolumns{6} 
\tablewidth{0pc}
\tablecaption{Details of CHAZSS {\it Chandra} and {\it RXTE} observations}
\tablehead{Object & Observation & {\it Chandra} & {\it Chandra} & {\it RXTE} & {\it RXTE}\\
 & Date & ObsID & Exposure (ks) & ObsID (ks) & Good time (ks)
 }
\startdata
4U~1636$-$53 & 2006 Mar 22  & 6635 & 23.2 & 91152-05-01 & 17.6  \\
4U~1636$-$53 & 2006 July 2  & 6636 & 25.2 & 91152-05-02 & 19.4  \\
4U~1735$-$44 & 2006 Aug 17  & 6637 & 24.2 & 91152-06-01 & 13.8  \\
4U~1735$-$44 & 2007 Mar 15  & 6638 & 25.1 & 91152-06-02 & 11.2  \\
GX~17+2      & 2006 May 10  & 6629 & 23.7 & 91152-02-01 & 13.4  \\
GX~17+2      & 2006 Aug 19  & 6630 & 24.1 & 91152-02-02 & 11.9  \\
4U~1820$-$30 & 2006 Aug 12  & 6633 & 25.2 & 91152-04-01 & 13.2  \\
4U~1820$-$30 & 2006 Oct 20  & 6634 & 25.1 & 91152-04-02 & 14.0  \\
GX~340+0     & 2006 May 19  & 6631 & 25.1 & 91152-03-01 & 18.8  \\
GX~340+0     & 2006 June 21 & 6632 & 23.7 & 91152-03-02 & 13.3  \\
GX~349+2     & 2006 July 4  & 6628 & 12.6 & 91152-01-02 & 6.6   \\
GX~349+2     & 2006 Aug 20  & 7336 & 12.1 & 91152-01-03 & 9.4
\enddata
\label{tab:obs}
\tablecomments{All {\it Chandra} observations were performed with the HETGS/ACIS instrument.}
\end{deluxetable}

\clearpage
\begin{turnpage}
\begin{deluxetable}{ccccccccccccc}
\tabletypesize{\small}
\tablecolumns{13} 
\tablewidth{0pc}
\tablecaption{Best fitting continuum spectral parameters for the hard state}
\tablehead{Object & Observation & $N_H$ (CXO) & $N_H$ (RXTE) & \multicolumn{2}{c}{Blackbody} & \multicolumn{4}{c}{Broken Power-law} & 2-10 keV flux & 10-20 keV flux & $\chi_\nu^2$ \\
 & & ($10^{22}$ cm$^{-2}$) & ($10^{22}$ cm$^{-2}$) & kT (keV) & Norm. ($10^{-2}$)\tablenotemark{a} & $\Gamma_1$ & $E_{br}$ (keV) & $\Gamma_2$ & Norm.\tablenotemark{b} & ($10^{-9}$ erg cm$^{-2}$ s$^{-1}$) & ($10^{-9}$ erg cm$^{-2}$ s$^{-1}$) &}
\startdata
4U~1636$-$53 & 1 & $0.57\pm0.19$ & $1.83\pm0.42 $ & $1.23\pm0.05$ & $0.25\pm0.05$ & $1.70\pm0.05$ & $17.1\pm1.6$  & $1.97\pm0.07$ & $0.24\pm0.03$ & $1.1\pm0.1$   & $0.6\pm0.1$  & 0.80    
\enddata
\tablecomments{The parameters are from joint spectral fits to the {\it Chandra} HEG and {\it RXTE} PCA spectra.  In the joint fitting all continuum parameters were tied except for $N_H$. All uncertainties are quoted at the 90\% confidence level.}
\tablenotetext{a}{Normalization of the blackbody component is defined as (L/$10^{39}$ erg s$^{-1}$)/(D/10 kpc)$^2$}
\tablenotetext{b}{Normalization of the broken power-law is defined as photons keV$^{-1}$ cm$^{-2}$ s$^{-1}$ at 1 keV}
\label{tab:spectra_hard}
\end{deluxetable}

\begin{deluxetable}{ccccccccccccc}
\tabletypesize{\small}
\tablecolumns{13} 
\tablewidth{0pc}
\tablecaption{Best fitting continuum spectral parameters for soft/transitional states}
\tablehead{Object & Obs. & $N_H$ (CXO) & $N_H$ (RXTE) & \multicolumn{2}{c}{Blackbody} & \multicolumn{2}{c}{Disk Blackbody} & \multicolumn{2}{c}{Power-law} & 2-10 keV flux & 10-20 keV flux & $\chi_\nu^2$ \\
 & & ($10^{22}$ cm$^{-2}$) & ($10^{22}$ cm$^{-2}$) & kT (keV) & Norm. ($10^{-2}$)\tablenotemark{a} & kT (keV) & Norm.\tablenotemark{b} & $\Gamma$ & Norm.\tablenotemark{c} & ($10^{-9}$ erg cm$^{-2}$ s$^{-1}$) & ($10^{-9}$ erg cm$^{-2}$ s$^{-1}$) & }
\startdata
4U~1636$-$53 & 2 & $0.39\pm0.12$ & $1.51\pm0.32$ & $2.20\pm0.06$ & $1.3\pm0.1$ & $1.21\pm0.04$ & $50.3\pm6.7$  & $2.50^{+0.0}_{-0.16}$ & $0.62\pm0.13$ & $2.7\pm0.1$   & $0.5\pm0.1$  & 0.86\\
4U~1735$-$44 & 1 & 0.30 (fixed)  & $1.38\pm0.19$ & $2.46\pm0.02$ & $2.1\pm0.1$ & $1.52\pm0.02$ & $24.7\pm1.4$  & $2.50^{+0.0}_{-0.05}$ & $0.40\pm0.04$ & $3.2\pm0.3$   & $0.8\pm0.1$  & 0.84\\
	     & 2 & 0.30 (fixed)  & $1.06\pm0.18$ & $2.59\pm0.03$ & $2.7\pm0.1$ & $1.74\pm0.03$ & $18.4\pm1.8$  & $2.50^{+0.0}_{-0.07}$ & $0.29\pm0.04$ & $3.9\pm0.3$   & $1.1\pm0.1$  & 0.92\\
GX~17+2      & 1 & $3.18\pm0.03$ & $4.63\pm0.08$ & $2.66\pm0.04$ & $7.3\pm0.1$ & $1.57\pm0.01$ & $152.5\pm5.7$ & --                    & --          & $12.1\pm 0.9$   & $2.9\pm0.1$  & 1.02\\
             & 2 & $3.12\pm0.05$ & $3.67\pm0.14$ & $2.54\pm0.02$ & $4.4\pm0.1$ & $1.55\pm0.03$ & $152.2\pm6.8$ & --                    & --          & $10.5\pm0.8$    & $1.8\pm0.1$  & 0.92\\
4U~1820$-$30 & 1 & 0.15 (fixed)  & $1.74\pm0.33$ & $2.36\pm0.03$ & $4.4\pm0.2$ & $1.39\pm0.05$ & $62.1\pm9.5$  & $2.17\pm0.20$         & $0.21\pm0.10$ & $5.6\pm0.6$   & $1.4\pm 0.1$ & 0.91\\
             & 2 & 0.15 (fixed)  & $1.82\pm0.27$ & $2.41\pm0.03$ & $6.0\pm0.2$ & $1.47\pm0.05$ & $73.9\pm6.4$  & $2.08\pm0.19$         & $0.18\pm0.07$ & $7.9\pm0.7$   & $2.0\pm0.2$  & 1.24\\
GX~340+0     & 1 & $6.41\pm0.05$ & $6.25\pm0.26$ & $2.66\pm0.04$ & $2.4\pm0.1$ & $1.63\pm0.01$ & $130.0\pm4.5$ & --                    & --          & $8.2\pm0.6$     & $1.3\pm0.1$  & 0.90\\
             & 2 & $6.47\pm0.05$ & $6.29\pm0.28$ & $2.54\pm0.03$ & $2.9\pm0.3$ & $1.59\pm0.01$ & $139.3\pm5.2$ & --                    & --          & $8.1\pm0.6$     & $1.4\pm0.1$  & 1.05\\        
GX~349+2     & 1 & $1.79\pm0.06$ & $3.13\pm0.14$ & $2.57\pm0.01$ & $7.2\pm0.1$ & $1.86\pm0.02$ & $88.4\pm1.1$  & --                    & --          & $15.9\pm1.3$    & $3.4\pm0.3$  & 0.96\\
             & 2 & $1.53\pm0.02$ & $2.59\pm0.30$ & $2.46\pm0.03$ & $5.8\pm0.3$ & $1.67\pm0.03$ & $105.1\pm5.4$ & --                    & --          & $12.5\pm1.2$    & $2.3\pm0.2$  & 0.96	     
\enddata
\tablecomments{The parameters are from joint spectral fits to the {\it Chandra} HEG and {\it RXTE} PCA spectra.  In the joint fitting all continuum parameters were tied except for $N_H$. All uncertainties are quoted at the 90\% confidence level. For GX~17+2 and GX~349+2 the fits to the {\it Chandra} spectra also include a Gaussian to model the iron emission line.  Best-fitting parameters for these Gaussian are given in Table \ref{tab:lines}.}
\tablenotetext{a}{Normalization of the blackbody component is defined as (L/$10^{39}$ erg s$^{-1}$)/(D/10 kpc)$^2$}
\tablenotetext{b}{Normalization of the disk blackbody is defined as (R$_{in}$/km)/(D/10 kpc)$^2$ $\times \cos{\theta}$, where R$_{in}$ is the inner disk radius and $\theta$ the angle of the disk.}
\tablenotetext{c}{Normalization of the power-law is defined as photons keV$^{-1}$ cm$^{-2}$ s$^{-1}$ at 1 keV}
\label{tab:spectra}
\end{deluxetable}

\end{turnpage}
\clearpage
\begin{deluxetable}{ccccccc}
\tabletypesize{\small}
\tablecolumns{7} 
\tablewidth{0pc}
\tablecaption{Iron emission line properties from fits to {\it Chandra} spectra}
\tablehead{Object & Observation & Energy (keV) & FWHM (keV) & EW (eV) & Line flux (10$^{-3}$ photons cm$^{-2}$ s$^{-1}$) & Significance ($\sigma$)}
\startdata
GX~17+2      & 1 & $6.67\pm0.08$ & $0.31\pm0.12$ & $14.1\pm5.2$  & $2.0\pm0.7$ & 4.5\\
             & 2 & $6.70\pm0.06$ & $0.45\pm0.17$ & $33.0\pm10.3$ & $3.7\pm1.2$ & 5.3\\
GX~349+2     & 1 & $6.75\pm0.06$ & $0.26\pm0.14$ & $18.1\pm7.2$  & $3.2\pm1.3$ & 4.1\\
             & 2 & $6.71\pm0.06$ & $0.52\pm0.17$ & $53.0\pm14.4$ & $6.8\pm1.8$ & 6.1
\enddata
\tablecomments{All uncertainties are quoted at the 90\% confidence level. }
\label{tab:lines}
\end{deluxetable}

\begin{deluxetable}{cccccc}
\centering
\tablecolumns{6} 
\tablewidth{0pc}
\tablecaption{Upper limits on iron emission line non-detections from {\it Chandra} observations}
\tablehead{Object & Observation & \multicolumn{2}{c}{Assuming $\sigma=0.1$ keV} & \multicolumn{2}{c}{Assuming $\sigma=0.5$ keV}  \\
& & Energy (keV) & EW (eV) & Energy (keV) & EW (eV)}
\startdata
4U~1636$-$53 & 1 & 6.7 (fixed) & $<12.1$ & 6.7 (fixed)  & $<80.1$ \\
             & 2 & 6.7 (fixed) & $<8.1$  & 6.7 (fixed)  & $<31.9$ \\

4U~1735$-$44 & 1 & $6.76\pm0.11$ & $<26.0$ & 6.7 (fixed) & $<39.6$ \\
             & 2 & 6.7 (fixed)   & $<14.4$ & 6.7 (fixed) & $<29.4$ \\

4U~1820$-$30 & 1 & 6.7 (fixed) & $<8.5$ & 6.7 (fixed) & $<31.9$\\
             & 2 & 6.7 (fixed) & $<11.7$ & 6.7 (fixed) & $<40.6$\\

GX~340+0     & 1 & $6.66\pm0.19$ & $<11.2$ & 6.7 (fixed) & $<16.2$  \\
             & 2 & 6.7 (fixed)   & $<9.2$  & 6.7 (fixed) & $<8.2$  

\enddata
\tablecomments{Upper limits are quoted at the 95\% confidence level. }
\label{tab:nonsig_lines}
\end{deluxetable}

\end{document}